\documentclass[12pt,preprint]{aastex}
\usepackage{psfig,epsfig,amsfonts,amsmath,amssymb,graphicx,morefloats,appendix,tensor}
\usepackage{color}
\usepackage{natbib}
\usepackage{hyperref}
\begin{document}

\slugcomment{Accepted to ApJL: June 8, 2022}

\title{ALMA Images the Eccentric HD~53143 Debris Disk
}

\author{Meredith A. MacGregor\altaffilmark{1}, Spencer A. Hurt\altaffilmark{1}, Christopher C. Stark\altaffilmark{2}, Ward S. Howard\altaffilmark{1}, Alycia J. Weinberger\altaffilmark{3}, Bin Ren\altaffilmark{4, 5}, Glenn Schneider\altaffilmark{6}, Elodie Choquet\altaffilmark{7}, Dmitri Mawet\altaffilmark{8, 9}}

\altaffiltext{1}{Department of Astrophysical and Planetary Sciences, University of Colorado, 2000 Colorado Avenue, Boulder, CO 80309, USA}
\altaffiltext{2}{NASA Goddard Space Flight Center, Exoplanets and Stellar Astrophysics Laboratory, Greenbelt, MD 20771, USA}
\altaffiltext{3}{Earth \& Planets Laboratory, Carnegie Institution for Science, 5241 Broad Branch Road NW, Washington, DC 20015, USA}
\altaffiltext{4}{Universit\'{e} Grenoble Alpes, Institut de Plan\'{e}tologie et d'Astrophysique (IPAG), F-38000 Grenoble, France} \altaffiltext{5}{Universit\'{e} C\^{o}te d'Azur, Observatoire de la C\^{o}te d'Azur, Lagrange, F-06304 Nice, France}
\altaffiltext{6}{Steward Observatory, The University of Arizona, 933 North Cherry Avenue, Tucson, AZ 85721, USA}
\altaffiltext{7}{Aix Marseille Univ, CNRS, CNES, LAM, Marseille, France}
\altaffiltext{8}{Department of Astronomy, California Institute of Technology, MC 249-17, 1200 East California Boulevard, Pasadena, CA 91125, USA}
\altaffiltext{9}{Jet Propulsion Laboratory, California Institute of Technology, 4800 Oak Grove Drive, Pasadena, CA 91109, USA}

\begin{abstract}

We present ALMA 1.3 mm observations of the HD~53143 debris disk -- the first infrared or millimeter image produced of this $\sim1$~Gyr-old solar-analogue.  Previous HST STIS coronagraphic imaging did not detect flux along the minor axis of the disk which could suggest a face-on geometry with two ‘clumps’ of dust.  These ALMA observations reveal a disk with a strikingly different structure.  In order to fit models to the millimeter visibilities and constrain the uncertainties on the disk parameters, we adopt an MCMC approach. This is the most eccentric debris disk observed to date with a forced eccentricity of $0.21\pm0.02$, nearly twice that of the Fomalhaut debris disk, and also displays apocenter glow.  Although this eccentric model fits the outer debris disk well, there are significant interior residuals remaining that may suggest a possible edge-on inner disk, which remains unresolved in these observations.  Combined with the observed structure difference between HST and ALMA, these results suggest a potential previous scattering event or dynamical instability in this system.  We also note that the stellar flux changes considerably over the course of our observations, suggesting flaring at millimeter wavelengths.  Using simultaneous TESS observations, we determine the stellar rotation period to be $9.6\pm0.1$~days.

\end{abstract}


\section{Introduction}
\label{sec:intro}

While protoplanetary disks are the reservoirs for planet formation, debris disks are the fossil record.  Within these remnant disks, asteroids and comets (often referred to as `planetesimals') collide and grind to  replenish the population of small dust grains \cite[see review article by][]{Hughes:2018}.  The presence of debris disks around mature stars is often seen as an indication that earlier planet formation was successful in these systems \cite[e.g.,][]{Meshkat:2017}.  At later evolutionary stages, interior planets now sculpt and stir surrounding disks through gravitational perturbations.  Thus, observed structure in debris disks can be traced back to the dynamical influence of planets and used to place constraints on their mass and orbital properties \cite[e.g.,][and references therein]{Wyatt:1999,Hahn:2005,Deller:2005,Chiang:2009}.  Since the dominant exoplanet detection techniques (e.g., transits and radial velocities) miss analogues to the outer giant planets in our own Solar System \cite[e.g.,][]{Zhu:2021}, debris disk structure offers a tantalizing path forward to fill in gaps in the exoplanet census and help put our planetary system into context.

At 18.3~pc \citep{Gaia}, HD~53143 is a Sun-like \cite[G9V,][]{Torres:2006} star that gives us a glimpse into what a planetary system like our own could have looked like at an earlier stage in its evolution (age $\sim1$~Gyr, see discussion in Section~\ref{sec:star}).  The outer debris disk was first imaged in scattered light at visible wavelengths by \cite{Kalas:2006} using the Advanced Camera for Surveys (ACS) on the Hubble Space Telescope (HST), and subsequently imaged by \cite{Schneider:2014} using the HST Space Telescope Imaging Spectrograph (STIS).  These previous coronagraphic images did not detect flux along the minor axis of the disk, which could be interpreted as resulting from a face-on geometry with two `clumps' of dust that could be resonant structures.  At longer wavelengths, ISO observations indicated a high disk dust luminosity, ie. surface area, relative to the star (2$\times10^{-4}$) \citep{Moor:2006}.  Herschel PACS observations at 70 and 160~$\mu$m marginally resolved a bright ring with radius $\sim54$~AU and width $39$~AU that appeared elongated to the NW and SE \citep{Marshall:2021,Pearce:2022}.  However, due to Herschel’s poor spatial resolution the exact geometry of the disk remained unclear.

In this paper we present the first resolved image of the HD~53143 debris disk at either infrared or millimeter wavelengths made with the Atacama Large Millimeter/submillimeter Array (ALMA).  These observations show a striking eccentric disk with features that suggest a previous scattering event or dynamical instability.  We discuss the observations and modeling in Section~\ref{sec:obs} and \ref{sec:model_approach}, respectively.  Section~\ref{sec:model_results} presents the modeling results followed by a discussion in Section~\ref{sec:disc} and conclusions in Section~\ref{sec:conc}.

\section{Observations}
\label{sec:obs}

We used ALMA Band 6 centered at 1.36~mm (238~GHz) to observe the HD~53143 debris disk during Cycle 6 (2018.1.00461.S, PI Stark).  A total of eight scheduling blocks (SBs) were executed between 12--23 March 2019.  The details of these observations including the date, number of antennas, shortest and longest baseline lengths, total time on-source, precipitable water vapor (PWV, a measure of the weather quality), and resulting rms noise are provided in Table~\ref{tab:obs}.  

The goal of these observations was to resolve the dust structure of the HD~53143 debris disk, so the correlator was set-up to maximize sensitivity to the millimeter dust continuum emission.  Three spectral windows were devoted entirely to continuum emission with a bandwidth of 2~GHz each split into 128 channels and centered at frequencies of 231.5, 244, and 246~GHz.  A fourth spectral window with a reduced bandwidth of 1.875~GHz split into 3840 channels was centered on the $^{12}$CO J$=2-1$ line at 230.538~GHz, although no gas emission was detected.

The same calibration sources were used for all eight SBs.  Bandpass, flux, and atmospheric calibration along with pointing were performed using the bright blazar J0538-4405 ($21\fdg0$ away from the target).  Phase calibration made use of J0700-6610.  All data processing, calibration, and imaging commands were executed in \texttt{CASA} \cite[version 5.4.0, ][]{McMullin:2007}.  Calibrated visibilities produced by the ALMA pipeline were then time-averaged in 30~sec intervals to reduce data volume and improve modeling speed.

\section{Results and Analysis}
\label{sec:results}

\begin{figure}[t]
  \begin{center}
       \includegraphics[scale=0.9]{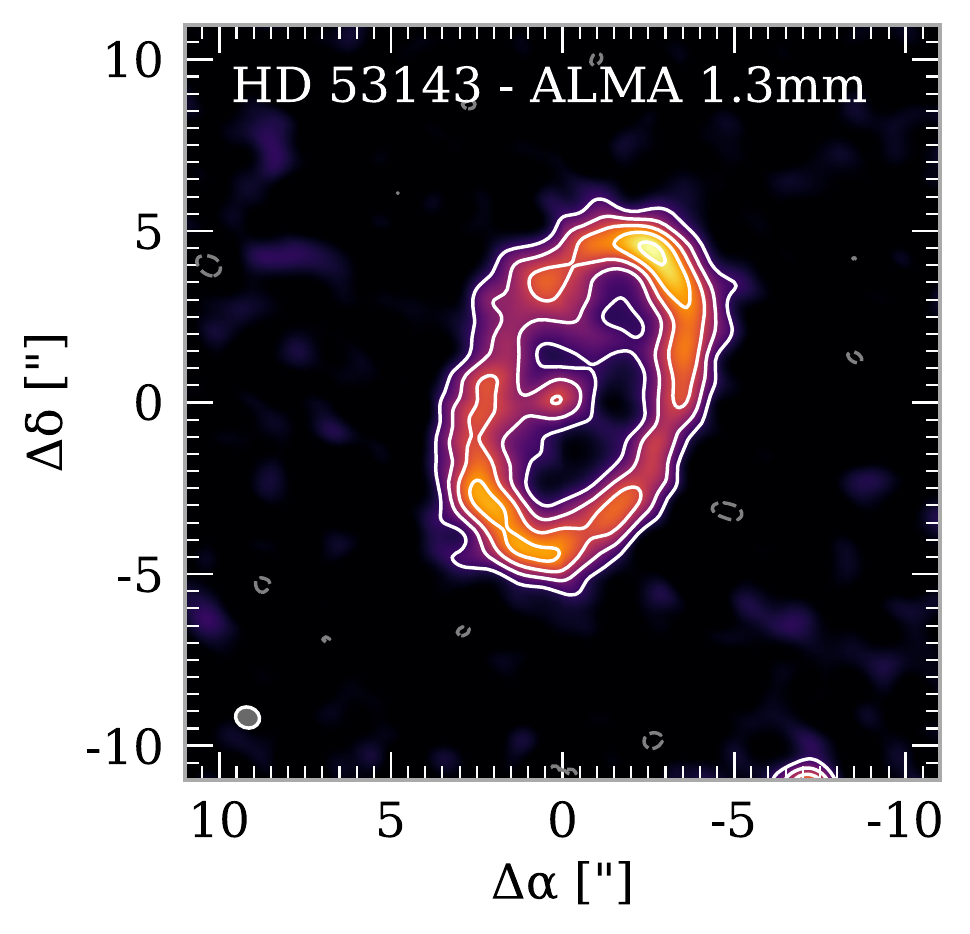}
  \end{center}
\caption{\small The first millimeter image of HD~53143 taken with ALMA shows an asymmetric distribution of dust with a bright point source in the center coincident with the star.  Contours are in steps of $3\sigma$ ($3\times$ the rms noise of 5.6~$\mu$Jy~beam$^{-1}$).  The gray ellipse in the lower left corner indicates the $1\farcs13\times0\farcs92$ synthesized beam with natural weighting.
}
\label{fig:fig1}
\end{figure}

Our new ALMA image of HD~53143 is shown in Figure~\ref{fig:fig1}.  This is the first time that the millimeter dust continuum emission has been imaged for this debris disk system.  The disk is noticeably asymmetric.  The bright point source at the image center is coincident ($<0\farcs08$ offset constrained by the modeling described in Sections~\ref{sec:model_approach} and \ref{sec:model_results} below) with the stellar position (RA = 06:59:59.2, DEC = -61:20:05.2), but offset from the disk center.  A brightness difference is also apparent between the two `ansae' (the limb-brightened edges of the disk), with the northwest side farther from the star roughly $3\sigma$ brighter than the southeast side closer to the star.    

\subsection{Modeling Approach}
\label{sec:model_approach}

In order to quantify the significance of the brightness asymmetry of the HD~53143 disk, we fit both symmetric and eccentric disk models to the ALMA data.  We follow the modeling approach first described in \cite{MacGregor:2013} and updated for eccentric disks in \cite{MacGregor:2017} in order to effectively constrain uncertainties and efficiently explore parameter space.  Visibilities are sampled from models of the sky brightness using the \texttt{galario} package \citep{Tazzari:2018} and compared directly to the ALMA visibilities within a Markov Chain Monte Carlo (MCMC) framework using the \texttt{emcee} package \citep{Foreman-Mackey:2013}.  For the symmetric model, we assume that the disk is radially symmetric and the surface brightness is described by a power law between an inner radius $R_\text{in}$ and an outer radius $R_\text{out}$: $I_\nu\propto r^{-0.5}$, where the power-law index describes the temperature profile and the surface density is assumed to be constant as a function of radius.  

For the eccentric model, we first populate the complex eccentricity space defined by the forced eccentricity and argument of periastron ($e_f$ and $\omega_f$, respectively) and the proper eccentricity and argument of periastron ($e_p$ and $\omega_p$, respectively) following \citep{Wyatt:1999}.  While $\omega_p$ is assumed to be randomly distributed from 0 to $2\pi$, $e_f$, $e_p$, and $\omega_f$ are all left as free parameters. We then compute the true anomaly, $f$, using the \texttt{kepler} package \citep{exoplanet:joss}, calculate the radial orbital position for each particle, and create a two-dimensional image by binning using a desired pixel scale (0.8~AU).  By assuming the same $r^{-0.5}$ temperature profile as for the symmetric model, we can then determine the flux in each pixel.  In order to sample the eccentricity parameter space completely and avoid the impact of shot noise where too few particles can make model images with identical parameters have different $\chi^2$ values \citep{Kennedy:2020}, we use a minimum of $10^7$ particles.  \cite{Kennedy:2020} also include a proper eccentricity dispersion in their model of the Fomalhaut disk to try to account for the fact that the apocenter side of the disk appears narrower than the pericenter side.  Since the HD~53143 disk does not appear to have this feature, we exclude this extra parameter in our models for simplicity.  Similarly, \cite{Lynch:2022} allow the forced eccentricity to vary with semi-major axis, which we also do not include here but could impact the derived parameter constraints.  

In both models, the vertical density profile of the disk as a function of radius is Gaussian and the standard deviation is given by the scale height $H(r) = hr$.  Given the resolution of the data, we assume that the aspect ratio $h$ is constant and fit for it as a free parameter.  Both models are also normalized to a total flux $F_\text{disk} = \int I_\nu d\Omega$ and include a point source to describe the detected stellar flux, $F_\text{star}$, and offsets relative to the pointing position in RA and DEC, $\Delta \alpha$ and $\Delta \delta$, respectively.

We assume uniform priors for all parameters, although some bounds are applied in order to ensure that the models are physical: $F_\text{disk}>0$, $F_\text{star}>0$, $0<R_\text{in}<R_\text{out}$.  To fully explore the parameter space of each model and ensure that all parameters have converged, we use $\sim10^6$ iterations (100 walkers, 11,450 and 12,750 steps for the eccentric and symmetric models, respectively ).  We consider the MCMC converged once the number of steps is more than 50 times greater than the autocorrelation function for each parameter.  The one-dimensional marginalized probability distributions for all model parameters appear Gaussian.  For the eccentric model, we do note a degeneracy between the forced ($e_f$, global orbital shape of the disk) and proper ($e_p$, intrinsic orbital scatter for individual particles) eccentricity.

\subsection{Modeling Results}
\label{sec:model_results}

\begin{figure}[t]
  \begin{center}
       \includegraphics[scale=0.55]{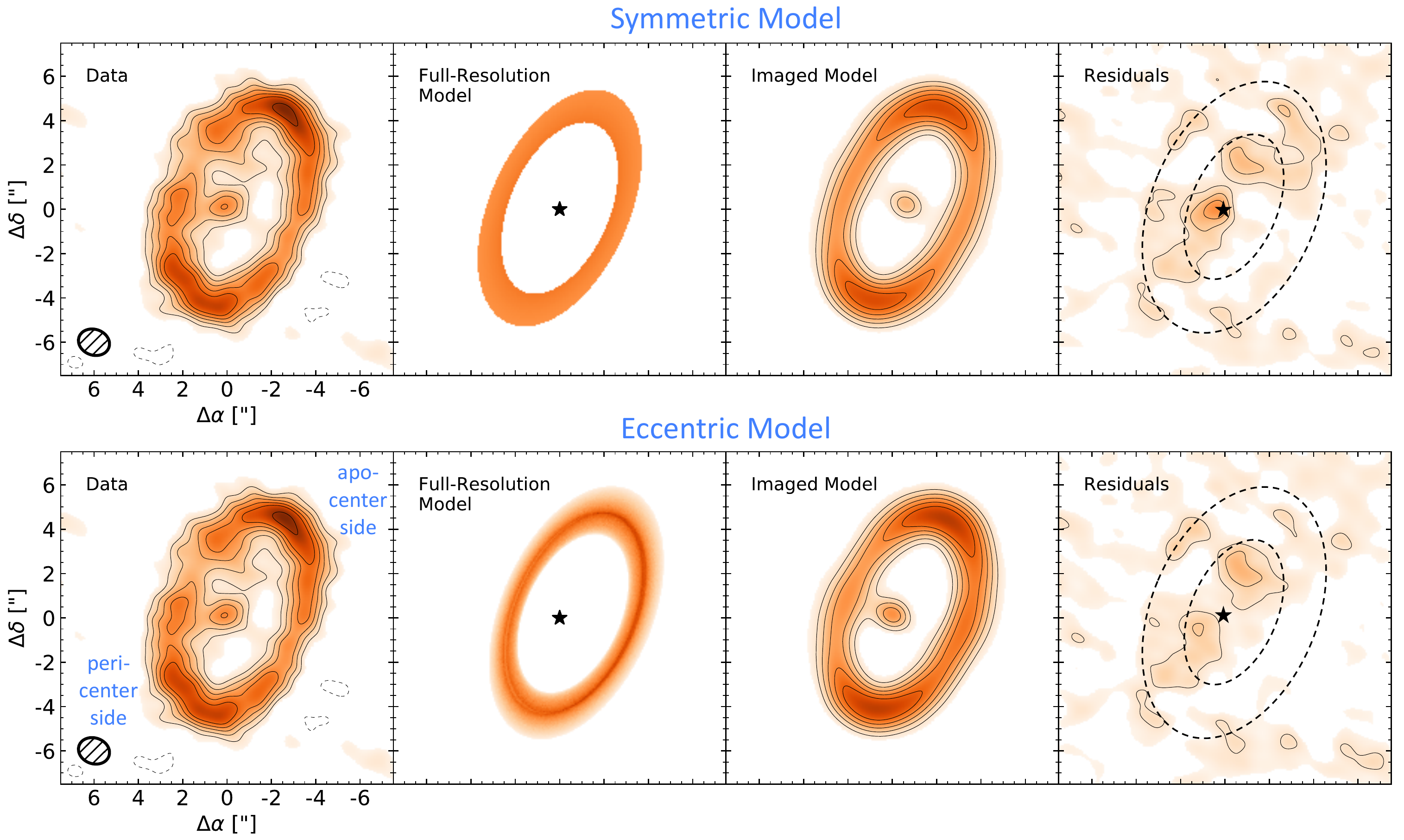}
  \end{center}
\caption{\small An eccentric disk model \emph{(bottom)} provides a better fit to the ALMA data than a symmetric one \emph{(top)}. In both figures, panels are as follows: \emph{(left)} the ALMA 1.3 mm continuum image, \emph{(left, center)} the best-fit model at full resolution, \emph{(right, center)} the best-fit model imaged like the data with no noise, and \emph{(right)} the residuals produced by subtracting the best-fit model from the data.  For reference, the dashed outline and star symbol indicate the edges of the outer disk and the stellar position in the residual image.  Contours are in steps of $3\sigma$ ($3\times$ the rms noise of 5.6~$\mu$Jy~beam$^{-1}$) in all panels.  As in Figure~\ref{fig:fig1}, the gray ellipse in the lower left corner indicates the $1\farcs13\times0\farcs92$ synthesized beam with natural weighting.
}
\label{fig:fig2}
\end{figure}

Table~\ref{tab:results} presents the best-fit model parameters for both the symmetric and eccentric models.  The symmetric model is unable to reproduce the data without significant stellar offsets, which makes it clear that an eccentric model is required to describe the HD~53143 debris disk.  Figure~\ref{fig:fig2} shows the ALMA 1.3 mm data (left panel) along with the best-fit symmetric and eccentric models (top and bottom, respectively) displayed at full resolution (left center) and imaged like the ALMA data (right center).  The residuals resulting from subtracting the best-fit model from the data are shown at right.  It is clear `by eye' that the eccentric model does a much better job fitting the structure of the outer debris disk.  The symmetric model leaves $>3\sigma$ residuals on the apocenter (northwest) side of the disk.   Both models leave $>6\sigma$ residuals interior to the disk.  These significant residuals are clear evidence that there is additional structure in this system that we are currently failing to resolve with our observations.

\section{Discussion}
\label{sec:disc}

We have presented the first millimeter image of the HD~53143 debris disk.  This new ALMA image shows that the disk is clearly eccentric exhibiting the same apocenter glow seen previously in the Fomalhaut debris disk.  Here, we discuss the eccentricity of the disk, a potential inner disk, and the dynamical implications of these results.  In addition, we compare the structure of the disk as observed by ALMA and HST and consider apparent variability in the star's emission during the course of our observations that could be due to flaring.

\subsection{Eccentric Structure}
\label{sec:eccentric}

HD~53143 is the most eccentric debris disk observed to date ($e_f=0.21\pm0.02$)  and the second example of a disk in which we have detected apocenter glow.  The first conclusive detection of this effect was made using ALMA observations of the Fomalhaut debris disk \cite[$e_f=0.12\pm0.01$,][]{MacGregor:2017}.  These new observations of HD~53143 show that this effect can be expected in any eccentric debris disk.  Since bodies on Keplerian orbits move more slowly at apocenter than pericenter, theory predicts that we should see a surface density enhancement at this location in the disk \citep{Pan:2016}.  At shorter infrared wavelengths, grain temperature dominates producing a brightness enhancement at pericenter closer to the star.  At longer millimeter wavelengths, observations are more sensitive to surface density and apocenter glow can be seen.  \cite{Lynch:2022} argue that the interpretation from \cite{Pan:2016} fails to account for the greater area over which dust is spread at apocenter, and that debris disks might exhibit either pericenter or apocenter glow at longer wavelengths depending on the resolution of the observations.  We note, however, that the spread in orbits at apocenter vs. pericenter also depends on the free eccentricity, which is not included in these new models so further investigation is needed to fully explore this effect. 

 Unlike Fomalhaut, however, HD~53143 appears to have a high proper eccentricity with $e_p = 0.11\pm0.01$ compared to $e_p=0.06\pm0.04$ \citep{MacGregor:2017}.  The proper eccentricity contributes to the overall width of the disk, so the high best-fit $e_p$ of the HD~53143 debris disk could have interesting implications for the structure and dynamics of this system. Future observations with higher angular resolution would better resolve the disk width, confirm this result, and distinguish between different potential sculpting scenarios by searching for azimuthal variation in the disk width.  Low $e_p$ ($e_p << e_f$, not likely for this system but not completely ruled out by the current observations) generally indicates that particles are on apsidally aligned orbits (i.e., they share the same pericenter position) resulting in a disk that is narrower at pericenter than at apocenter.  This `nesting' of orbits is frequently seen in other eccentric planetary systems.  For example, the `hot classical' Kuiper Belt Objects (KBOs) in our own solar system share a pericenter position determined by resonances with Neptune but with varying eccentricities \cite[e.g.,][]{Malhotra:2019}, and apsidal alignment is apparent in the Upsilon Andromeda planetary system potentially mediated by a circumstellar disk \citep{Chiang:2001}.  For HD~53143, our current observations support a scenario where the proper eccentricity is comparable to the forced eccentricity ($e_p \sim e_f$).  In this case, the width should be more uniform azimuthally and higher resolution observations would not see a difference between apocenter and pericenter.  A more extreme and varied ratio of widths between apocenter and pericenter is possible if $e_p > e_f$. \cite{Dermott:1980} propose a combination of self-gravitation, particle collisions, and close-packing to produce an extreme ‘pinching’ of a belt at pericenter. The $\epsilon$ ring of Uranus has a width of $\sim20$~km at pericenter compared to $\sim96$~km at apocenter, implying a width ratio of $\sim1:5$ \citep{Elliot:1984}. 

HD~53143 is also notable when compared with other eccentric disk because of its width.  Our model fits yield $\Delta a = 19.7\pm2.5$~AU, which gives $\Delta a/a \sim 0.22$.  Other eccentric debris disks are significantly narrower with $\Delta a/a$ values of $\sim 0.10$ for Fomalhaut \citep{MacGregor:2017} and $\sim 0.14$ for HD~202628 \citep{Faramaz:2019}.  Secular perturbations from an eccentric planet predict that the width of an eccentric debris disk should be $2ae_f$, and both Fomalhaut and HD~202628 are far narrower than this \citep{Kennedy:2020}.  Even though HD~53143 is comparably broader, it is still narrower than the theoretical prediction of $\sim40$~AU given the best-fit semi-major axis and forced eccentricity.  \cite{Kennedy:2020} propose that narrower debris disks could be created if earlier planet perturbations during the gas-rich protoplanetary disk phase put planetesimals on initially eccentric orbits.  This is an intriguing possibility to consider for the HD~53143 system given the presence of a potentially misaligned inner disk (Section~\ref{sec:innerdisk}) and vertically scattered small grains (Section~\ref{sec:comparison}), which both also suggest a previous scattering event or dynamical instability.

\subsection{Inner Dust Emission?}
\label{sec:innerdisk}

Given the significant excess of millimeter emission interior to the outer debris ring, it is clear that there is a source of dust in the inner part of the HD~53143 system.  The two $6\sigma$ peaks on either side of the star are suggestive of limb brightening from an inclined inner disk.  From the image, we can estimate the potential radius and width of this inner ring to be $\sim25$~AU ($1\farcs4$) and $\sim~5$~AU ($0\farcs3$), respectively, given the separation of the two peaks, with a maximum flux density of $\sim80$~$\mu$Jy.  We attempted to include this inner component in our eccentric model, but results were limited by the data quality.  By fixing the disk flux and radial position, we were able to return a fit on the inclination to be $85\degr^{+5}_{-15}$, close to edge-on.  However, since the current ALMA observations only achieve a resolution of roughly 20~AU, this potential inner disk is not well-resolved and the geometry is not well-constrained.  Earlier HST observations detected a circularly symmetric excess, indicating the presence of an inner ring.  The inferred geometry is quite different, however, with the inner disk extending from 5 -- 55~AU ($0\farcs3-3\farcs1$) and viewed face-on \cite[i.e., inclination of $0\degr$,][]{Schneider:2014}.  Interestingly, either a face-on or close to edge-on geometry would imply that the inner and outer rings are not aligned.  

Another possibility is that dust from the outer ring is being passed inwards.  The $\eta$ Corvi system includes a hot inner dust ring and a cold outer belt that has been previously resolved by ALMA.  In order to replenish the inner disk, material must flow from the outer to the inner disk.  \cite{Marino:2017} suggest that a stable planetary system could scatter material and feed a close-in collisional cascade.  They also discuss the possibility that grains could be scattered inwards by a planet and then transported through Poynting-Robertson (PR) drag.  PR drag generally affects small grains that do not emit efficiently at millimeter wavelengths.  However, an abundance of micron-sized grains can still produce emission at longer wavelengths as has been discussed previously for the extended halos observed towards HD~32297 and HD~61005 \citep{MacGregor:2018b}.  Future observations with higher spatial resolution are required to definitively show us what mechanism or structure is responsible for the observed inner excess emission in the HD~53143 system, and as a result reveal the dynamical history of this planetary system.

\subsection{Comparison to Scattered Light Observations}
\label{sec:comparison}

As discussed above, previous HST coronagraphic imaging did not detect flux along the minor axis of the disk \citep{Kalas:2006,Schneider:2014}, which could suggest that the HD~53143 disk might be face-on with resonant clumps of emission.   New HST STIS observations from program GO-16202 (Ren et al., in prep.) show that this interpretation is not correct and confirm that the disk is in fact inclined relative to the line of sight.  However, while the ALMA data is best fit by a thin ($h<0.04$) disk, the HST data suggest a significant scale height of $h\sim0.3$.  Unlike every other detected inclined debris disks, the HD~53143 disk shows no clear forward-scattering side in any of the previous HST images that matches the minor axis of the ALMA observation at visible wavelengths. In fact, it shows the opposite -- a likely absence of flux near the minor axes and an enhancement of flux near the ansae. Given the additional constraints from the ALMA observations, the two best physical explanations for this are a semi-inclined disk with either (1) a large scale height and isotropically scattering small grains or (2) a small scale height and density enhancements at the ansae.  ALMA observations trace thermal emission from roughly millimeter-sized grains.  HST observations trace scattered light from small, micron-sized grains.  Thus, if the modeled difference in scale height between these two observations is real, then the dynamics of large vs. small grains are different in the HD~53143 disk.  One possible explanation is that the small grains are getting stirred and puffed up by a giant planet orbiting within the disk \cite[e.g.,][]{Quillen:2006,Pan:2012}.  A significant scattering event, potentially involving the migration of giant planets, could also potentially puff up the disk, and might also help explain the potential misalignment between an inner and outer disk (see Section~\ref{sec:innerdisk}).

\subsection{Stellar Emission}
\label{sec:star}

\begin{figure}[t]
  \begin{center}
       \includegraphics[scale=0.7]{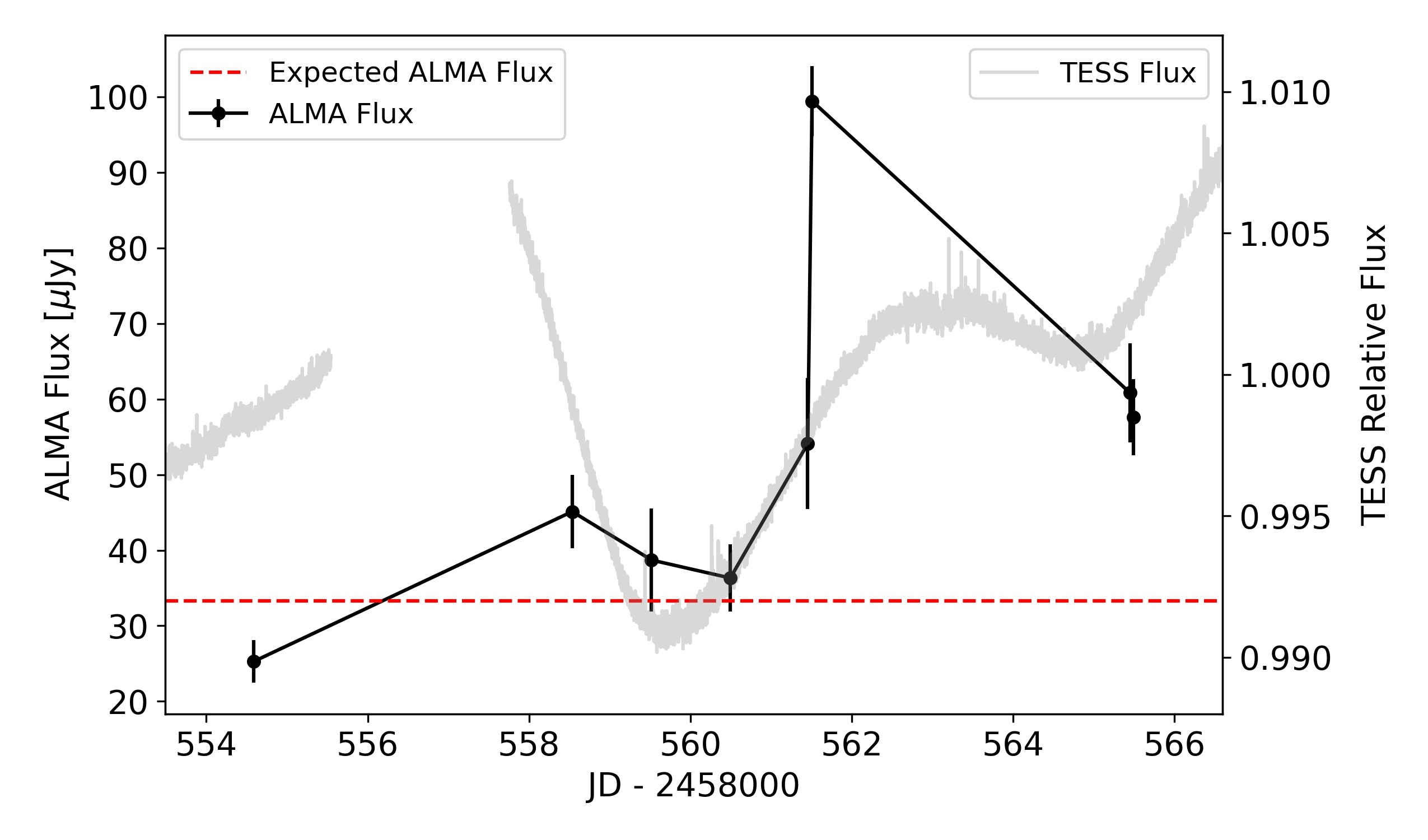}
  \end{center}
\caption{\small The flux density of the central star varies over the course of our eight ALMA observations, indicating potential flaring.  Each point in this light curve represents the best-fit stellar flux density in each individual observation.  For reference, the expected photospheric flux of 33~$\mu$Jy for a G9V star is indicated by the dashed red line.  Simultaneous TESS observations plotted in gray indicate significant spot modulation.
}
\label{fig:fig3}
\end{figure}

The ALMA observations of HD~53143 were executed over eight nights in March 2019.  Although the sensitivity is not high enough to split the data up further, we are able to fit the flux of the star in each of these eight observations separately.  The results are shown in Fig~\ref{fig:fig3}.  Assuming that HD~53143 is a G9V star \cite[temperature $\sim5413$~K, radius $\sim0.86$~$R_\odot$,][]{Fuhrmann:2015} the expected photospheric flux at 1.3~mm is 33~$\mu$Jy using a PHOENIX stellar atmosphere model \citep{Husser:2013}.  In observations 2 and 5--8, the best-fit flux for the star is significantly in excess of this value.  The excess is most extreme in observation 6, when the star is $>3\times$ brighter than we would expect in quiescence.  This transient behavior on roughly hour-long timescales is consistent with stellar flaring, which has been observed from several M dwarf stars \citep{MacGregor:2018a,MacGregor:2020}.  

Serendipitously, the Transiting Exoplanet Survey Satellite (TESS) observed HD~53143 simultaneously with our ALMA observations (light curve is shown in gray in Figure~\ref{fig:fig3}).  Although no significant flares are apparent, the periodic variations in brightness are indicative of spot modulation.  We expect that millimeter emission is similar to FUV in that it traces particle acceleration in flares, while TESS traces the resulting photospheric heating \citep{MacGregor:2020}. It is notable that stars observed to flare in the FUV (and therefore presumably in the millimeter) sometimes have no flares in the TESS bandpass \citep{Loyd:2020}.  Similarly, \cite{Williams:2014} suggest that radio-loud dwarfs may be explained by constant low-level particle acceleration without producing heating in other bands.  Because of its extreme southern declination, HD~53143 is actually near the continuous viewing zone for TESS and was observed in 14 additional sectors.  We use a Lomb-Scargle periodogram to determine the rotation period for the star to be $9.6\pm0.1$~days.  The rotation-age relation for a G9/K0 dwarf in \cite{Curtis:2019} gives 10~days as the expected rotation period at 1~Gyr, although some K0 dwarfs with 10~day periods are as young as 670~Myr.  This is now the best age determination for HD~53143.  The previous estimate was slightly higher at $1.1\pm0.13$~Gyr derived using ROSAT x-ray observations and the stellar activity-rotation relationship \citep{Pizzolato:2003}.


\section{Conclusions}
\label{sec:conc}

In this paper, we present the first ALMA image of the HD~53143 debris disk. This disk had only been imaged previously in scattered light with HST coronagraphic observations.  Our ALMA observations yield new insights on the structure of this young solar analogue, revealing a clearly eccentric disk that displays apocenter glow.  In order to constrain the geometry of the disk, we adopt an MCMC framework to fit models directly to the millimeter visibilities.  Our main conclusions are as follows:

\begin{enumerate}
    \item The best-fit model constrains the disk eccentricity to be $0.21\pm0.02$, considerably higher than the Fomalhaut debris disk.  In addition, the proper eccentricity of the disk is $0.11\pm0.01$, much higher than the Fomalhaut disk.  This result suggests that the orbits in the disk are not apsidally aligned with each other.
    \item Although this eccentric model reproduces the outer disk structure well, $>6\sigma$ residuals remain interior.  These could be due to an additional inner disk or dust from the outer ring being passed inwards.  
    \item Modeling of HST and ALMA observations does not produce consistent results for the structure and geometry of the outer disk.  Since HST observations trace micron-sized grains and ALMA observations trace millimeter-sized grains, this suggests that small and large grains are experiencing different sculpting mechanisms.
    \item The stellar flux varies considerably over the course of our observations, with a clear peak in observation 6.  We interpret this as evidence for millimeter flaring.  Simultaneous TESS observations suggest considerable spot modulation.  Using the complete TESS light curve, we have made the first determination of the rotation period for this star of $9.6\pm0.1$~days.
\end{enumerate}

The HD~53143 debris disk is an intriguing system that may have undergone a previous scattering event or dynamical instability.  Its unique characteristics certainly suggest a need for future study as a test bed to consider the effects of planetary migration on debris disk structure.  ALMA observations with higher resolution would be able to resolve an inner disk if present and look for any variation in the disk width as a function of azimuthal angle.  With this critical information, we would be able to place strong constraints on the orbits and masses of any planets in the system.

\vspace{0.8cm}
This paper makes use of the following ALMA data: ADS/JAO.ALMA \#2018.1.00461.S. ALMA is a partnership of ESO (representing its member states), NSF (USA) and NINS (Japan), together with NRC (Canada) and NSC and ASIAA (Taiwan) and KASI (Republic of Korea), in cooperation with the Republic of Chile. The Joint ALMA Observatory is operated by ESO, AUI/NRAO and NAOJ. The National Radio Astronomy Observatory is a facility of the National Science Foundation operated under cooperative agreement by Associated Universities, Inc.  The TESS data presented in this paper were obtained from the Mikulski Archive for Space Telescopes (MAST) at the Space Telescope Science Institute. The specific observations analyzed can be accessed via \dataset[10.17909/t9-nmc8-f686]{https://doi.org/10.17909/t9-nmc8-f686}.

M.A.M. acknowledges support for part of this research from the National Aeronautics and Space Administration (NASA) under award number 19-ICAR19\_2-0041.

\vspace{0.8cm}
\software{\texttt{CASA} \cite[v5.4.0][]{McMullin:2007}, \texttt{galario} \citep{Tazzari:2018}, \texttt{emcee} \citep{Foreman-Mackey:2013}, \texttt{kepler} \citep{exoplanet:joss}, \texttt{Lightkurve} \citep{lightkurve}, \texttt{astropy} \citep{astropy:2013,astropy:2018}, \texttt{astroquery} \citep{Ginsburg:2019}}

\pagebreak

\bibliography{References.bib}

\pagebreak

\begin{deluxetable}{l l l c c c c c}
\tablecolumns{8}
\tabcolsep0.1in\footnotesize
\tabletypesize{\small}
\tablewidth{0pt}
\tablecaption{2019 ALMA Observations of HD 53143 \label{tab:obs}}
\tablehead{
\colhead{Obs.} &
\colhead{Date} & 
\colhead{Time Range} &
\colhead{\# of} & 
\colhead{Baseline} &
\colhead{On-Source} &
\colhead{PWV} &
\colhead{rms}\\
\colhead{ID} &
\colhead{} & 
\colhead{(UTC)} &
\colhead{Ants.} & 
\colhead{Lengths (m)} &
\colhead{Time (min)} &
\colhead{(mm)} &
\colhead{(mJy)}
}
\startdata
1 & 12 March & 02:07:13.4 -- 03:04:03.2 & 46 & 14--360 & 49.63 & 1.29 & 14.6 \\
2 & 16 March & 00:50:56.3 -- 01:47:22.4 & 47 & 14--360 & 49.67 & 1.26 & 15.0 \\
3 & 17 March & 00:27:33.3 -- 01:24:01.0 & 47 & 14--360 & 49.62 & 1.74 & 16.7 \\
4 & 17--18 March & 23:53:09.7 -- 00:49:31.6 & 44 & 14--313 & 49.62 & 2.17 & 18.6 \\
5 & 18 March & 22:56:23.5 -- 23:53:00.0 & 45 & 14--313 & 49.63 & 3.08 & 20.8 \\
6 & 19 March & 00:13:01.4 -- 01:09:37.4 & 45 & 14--313 & 49.62 & 2.64 & 19.9 \\
7 & 22 March & 22:45:10.1 -- 23:41:39.2 & 47 & 14--360 & 49.65 & 2.18 & 16.3 \\
8 & 22--23 March & 23:53:32.8 -- 00:50:00.5 & 46 & 14--360 & 49.63 & 2.11 & 16.7 \\
\enddata
\end{deluxetable}

\begin{deluxetable}{l l l c c}
\tablecolumns{5}
\tabcolsep0.1in\footnotesize
\tabletypesize{\small}
\tablewidth{0pt}
\tablecaption{Best-Fit Model Parameters \label{tab:results}}
\tablehead{
\colhead{} &
\colhead{Parameter} &
\colhead{Description} & 
\colhead{Symmetric Model} &
\colhead{Eccentric Model}
}
\startdata
\emph{Disk Geometry} & $a$ & Disk radial position [au] & $88.3\pm0.9$ & $90.1\pm0.5$\\
 & $\Delta a$ & Disk width [au] & $29.7\pm1.0$ & $19.7\pm2.5$\\
 & $h$ & Disk scale height & $0.04\pm0.02$ & $0.04\pm0.02$ \\
 & $i$ & Inclination [$\deg$] & $56.1\pm0.4$ & $56.2\pm0.4$ \\
 & $PA$ & Position angle [$\deg$] & $156.4\pm0.4$ & $157.3\pm0.3$ \\
 & $\omega_f$ & Argument of pericenter [$\deg$] & -- & $112.8\pm2.1$ \\
 \hline
 \emph{Flux} & $F_\text{tot}$ & Total disk flux [mJy] & $1.36\pm0.03$ & $1.42\pm0.03$\\
 & $F_\text{star}$ & Total stellar flux [mJy] & $0.05\pm0.01$ & $0.05\pm0.01$\\
 \hline
 \emph{Eccentricity} & $e_f$ & Forced eccentricity & -- & $0.21\pm0.02$ \\
 & $e_p$ & Proper eccentricity & -- & $0.11\pm0.01$ \\
 \hline
 \emph{Offset} & $\Delta \alpha$ & RA offset [$\arcsec$] & $-0.60\pm0.01$ & $0.07\pm0.06$ \\
 & $\Delta \delta$ & DEC offset [$\arcsec$] & $0.17\pm0.02$ & $0.04\pm0.04$
\enddata
\end{deluxetable}

\end{document}